\gdef\islinuxolivetti{F}
\gdef\PSfonts{T}
\magnification\magstep1

\newdimen\papwidth
\newdimen\papheight
\newskip\beforesectionskipamount  
\newskip\sectionskipamount 
\def\sectionskip{\vskip\sectionskipamount}
\def\beforesectionskip{\vskip\beforesectionskipamount}
\papwidth=16truecm
\if F\islinuxolivetti
\papheight=22truecm
\voffset=0.4truecm
\hoffset=0.4truecm
\else
\papheight=16truecm
\voffset=-1.5truecm
\hoffset=0.4truecm
\fi
\hsize=\papwidth
\vsize=\papheight
\nopagenumbers
\headline={\ifnum\pageno>1 {\hss\tenrm-\ \folio\ -\hss} \else
{\hfill}\fi}
\newdimen\texpscorrection
\texpscorrection=0.15truecm 

\def\sectionsize{\twelvepoint}
\def\sectiontype{\bf}
\def\subsectionsize{}
\def\subsectiontype{\bf}
\def\em{\sl}
\newfam\truecmsy
\newfam\truecmr
\newfam\msbfam
\newfam\scriptfam
\newfam\truecmsy
\newskip\ttglue 
\if T\islinuxolivetti
\papheight=11.5truecm
\fi
\if F\PSfonts
\font\twelverm=cmr12
\font\tenrm=cmr10
\font\eightrm=cmr8
\font\sevenrm=cmr7
\font\sixrm=cmr6
\font\fiverm=cmr5

\font\twelvebf=cmbx12
\font\tenbf=cmbx10
\font\eightbf=cmbx8
\font\sevenbf=cmbx7
\font\sixbf=cmbx6
\font\fivebf=cmbx5

\font\twelveit=cmti12
\font\tenit=cmti10
\font\eightit=cmti8
\font\sevenit=cmti7
\font\sixit=cmti6
\font\fiveit=cmti5

\font\twelvesl=cmsl12
\font\tensl=cmsl10
\font\eightsl=cmsl8
\font\sevensl=cmsl7
\font\sixsl=cmsl6
\font\fivesl=cmsl5

\font\twelvei=cmmi12
\font\teni=cmmi10
\font\eighti=cmmi8
\font\seveni=cmmi7
\font\sixi=cmmi6
\font\fivei=cmmi5

\font\twelvesy=cmsy10	at	12pt
\font\tensy=cmsy10
\font\eightsy=cmsy8
\font\sevensy=cmsy7
\font\sixsy=cmsy6
\font\fivesy=cmsy5
\font\twelvetruecmsy=cmsy10	at	12pt
\font\tentruecmsy=cmsy10
\font\eighttruecmsy=cmsy8
\font\seventruecmsy=cmsy7
\font\sixtruecmsy=cmsy6
\font\fivetruecmsy=cmsy5

\font\twelvetruecmr=cmr12
\font\tentruecmr=cmr10
\font\eighttruecmr=cmr8
\font\seventruecmr=cmr7
\font\sixtruecmr=cmr6
\font\fivetruecmr=cmr5

\font\twelvebf=cmbx12
\font\tenbf=cmbx10
\font\eightbf=cmbx8
\font\sevenbf=cmbx7
\font\sixbf=cmbx6
\font\fivebf=cmbx5

\font\twelvett=cmtt12
\font\tentt=cmtt10
\font\eighttt=cmtt8

\font\twelveex=cmex10	at	12pt
\font\tenex=cmex10

\font\twelvemsb=msbm10	at	12pt
\font\tenmsb=msbm10
\font\eightmsb=msbm8
\font\sevenmsb=msbm7
\font\sixmsb=msbm6
\font\fivemsb=msbm5

\font\twelvescr=eusm10 at 12pt
\font\tenscr=eusm10
\font\eightscr=eusm8
\font\sevenscr=eusm7
\font\sixscr=eusm6
\font\fivescr=eusm5
\fi
\if T\PSfonts
\font\twelverm=ptmr	at	12pt
\font\tenrm=ptmr	at	10pt
\font\eightrm=ptmr	at	8pt
\font\sevenrm=ptmr	at	7pt
\font\sixrm=ptmr	at	6pt
\font\fiverm=ptmr	at	5pt

\font\twelvebf=ptmb	at	12pt
\font\tenbf=ptmb	at	10pt
\font\eightbf=ptmb	at	8pt
\font\sevenbf=ptmb	at	7pt
\font\sixbf=ptmb	at	6pt
\font\fivebf=ptmb	at	5pt

\font\twelveit=ptmri	at	12pt
\font\tenit=ptmri	at	10pt
\font\eightit=ptmri	at	8pt
\font\sevenit=ptmri	at	7pt
\font\sixit=ptmri	at	6pt
\font\fiveit=ptmri	at	5pt

\font\twelvesl=ptmro	at	12pt
\font\tensl=ptmro	at	10pt
\font\eightsl=ptmro	at	8pt
\font\sevensl=ptmro	at	7pt
\font\sixsl=ptmro	at	6pt
\font\fivesl=ptmro	at	5pt

\font\twelvei=cmmi12
\font\teni=cmmi10
\font\eighti=cmmi8
\font\seveni=cmmi7
\font\sixi=cmmi6
\font\fivei=cmmi5

\font\twelvesy=cmsy10	at	12pt
\font\tensy=cmsy10
\font\eightsy=cmsy8
\font\sevensy=cmsy7
\font\sixsy=cmsy6
\font\fivesy=cmsy5
\font\twelvetruecmsy=cmsy10	at	12pt
\font\tentruecmsy=cmsy10
\font\eighttruecmsy=cmsy8
\font\seventruecmsy=cmsy7
\font\sixtruecmsy=cmsy6
\font\fivetruecmsy=cmsy5

\font\twelvetruecmr=cmr12
\font\tentruecmr=cmr10
\font\eighttruecmr=cmr8
\font\seventruecmr=cmr7
\font\sixtruecmr=cmr6
\font\fivetruecmr=cmr5

\font\twelvebf=cmbx12
\font\tenbf=cmbx10
\font\eightbf=cmbx8
\font\sevenbf=cmbx7
\font\sixbf=cmbx6
\font\fivebf=cmbx5

\font\twelvett=cmtt12
\font\tentt=cmtt10
\font\eighttt=cmtt8

\font\twelveex=cmex10	at	12pt
\font\tenex=cmex10

\font\twelvemsb=msbm10	at	12pt
\font\tenmsb=msbm10
\font\eightmsb=msbm8
\font\sevenmsb=msbm7
\font\sixmsb=msbm6
\font\fivemsb=msbm5

\font\twelvescr=eusm10 at 12pt
\font\tenscr=eusm10
\font\eightscr=eusm8
\font\sevenscr=eusm7
\font\sixscr=eusm6
\font\fivescr=eusm5
\fi
\def\eightpoint{\def\rm{\fam0\eightrm}%
\textfont0=\eightrm
  \scriptfont0=\sixrm
  \scriptscriptfont0=\fiverm 
\textfont1=\eighti
  \scriptfont1=\sixi
  \scriptscriptfont1=\fivei 
\textfont2=\eightsy
  \scriptfont2=\sixsy
  \scriptscriptfont2=\fivesy 
\textfont3=\tenex
  \scriptfont3=\tenex
  \scriptscriptfont3=\tenex 
\textfont\itfam=\eightit
  \scriptfont\itfam=\sixit
  \scriptscriptfont\itfam=\fiveit 
  \def\it{\fam\itfam\eightit}%
\textfont\slfam=\eightsl
  \scriptfont\slfam=\sixsl
  \scriptscriptfont\slfam=\fivesl 
  \def\sl{\fam\slfam\eightsl}%
\textfont\ttfam=\eighttt
  \def\tt{\fam\ttfam\eighttt}%
\textfont\bffam=\eightbf
  \scriptfont\bffam=\sixbf
  \scriptscriptfont\bffam=\fivebf
  \def\bf{\fam\bffam\eightbf}%
\textfont\scriptfam=\eightscr
  \scriptfont\scriptfam=\sixscr
  \scriptscriptfont\scriptfam=\fivescr
  \def\script{\fam\scriptfam\eightscr}%
\textfont\msbfam=\eightmsb
  \scriptfont\msbfam=\sixmsb
  \scriptscriptfont\msbfam=\fivemsb
  \def\bb{\fam\msbfam\eightmsb}%
\textfont\truecmr=\eighttruecmr
  \scriptfont\truecmr=\sixtruecmr
  \scriptscriptfont\truecmr=\fivetruecmr
  \def\truerm{\fam\truecmr\eighttruecmr}%
\textfont\truecmsy=\eighttruecmsy
  \scriptfont\truecmsy=\sixtruecmsy
  \scriptscriptfont\truecmsy=\fivetruecmsy
\tt \ttglue=.5em plus.25em minus.15em 
\normalbaselineskip=9pt
\setbox\strutbox=\hbox{\vrule height7pt depth2pt width0pt}%
\normalbaselines
\rm
}

\def\tenpoint{\def\rm{\fam0\tenrm}%
\textfont0=\tenrm
  \scriptfont0=\sevenrm
  \scriptscriptfont0=\fiverm 
\textfont1=\teni
  \scriptfont1=\seveni
  \scriptscriptfont1=\fivei 
\textfont2=\tensy
  \scriptfont2=\sevensy
  \scriptscriptfont2=\fivesy 
\textfont3=\tenex
  \scriptfont3=\tenex
  \scriptscriptfont3=\tenex 
\textfont\itfam=\tenit
  \scriptfont\itfam=\sevenit
  \scriptscriptfont\itfam=\fiveit 
  \def\it{\fam\itfam\tenit}%
\textfont\slfam=\tensl
  \scriptfont\slfam=\sevensl
  \scriptscriptfont\slfam=\fivesl 
  \def\sl{\fam\slfam\tensl}%
\textfont\ttfam=\tentt
  \def\tt{\fam\ttfam\tentt}%
\textfont\bffam=\tenbf
  \scriptfont\bffam=\sevenbf
  \scriptscriptfont\bffam=\fivebf
  \def\bf{\fam\bffam\tenbf}%
\textfont\scriptfam=\tenscr
  \scriptfont\scriptfam=\sevenscr
  \scriptscriptfont\scriptfam=\fivescr
  \def\script{\fam\scriptfam\tenscr}%
\textfont\msbfam=\tenmsb
  \scriptfont\msbfam=\sevenmsb
  \scriptscriptfont\msbfam=\fivemsb
  \def\bb{\fam\msbfam\tenmsb}%
\textfont\truecmr=\tentruecmr
  \scriptfont\truecmr=\seventruecmr
  \scriptscriptfont\truecmr=\fivetruecmr
  \def\truerm{\fam\truecmr\tentruecmr}%
\textfont\truecmsy=\tentruecmsy
  \scriptfont\truecmsy=\seventruecmsy
  \scriptscriptfont\truecmsy=\fivetruecmsy
\tt \ttglue=.5em plus.25em minus.15em 
\normalbaselineskip=12pt
\setbox\strutbox=\hbox{\vrule height8.5pt depth3.5pt width0pt}%
\normalbaselines
\rm
}

\def\twelvepoint{\def\rm{\fam0\twelverm}%
\textfont0=\twelverm
  \scriptfont0=\tenrm
  \scriptscriptfont0=\eightrm 
\textfont1=\twelvei
  \scriptfont1=\teni
  \scriptscriptfont1=\eighti 
\textfont2=\twelvesy
  \scriptfont2=\tensy
  \scriptscriptfont2=\eightsy 
\textfont3=\twelveex
  \scriptfont3=\twelveex
  \scriptscriptfont3=\twelveex 
\textfont\itfam=\twelveit
  \scriptfont\itfam=\tenit
  \scriptscriptfont\itfam=\eightit 
  \def\it{\fam\itfam\twelveit}%
\textfont\slfam=\twelvesl
  \scriptfont\slfam=\tensl
  \scriptscriptfont\slfam=\eightsl 
  \def\sl{\fam\slfam\twelvesl}%
\textfont\ttfam=\twelvett
  \def\tt{\fam\ttfam\twelvett}%
\textfont\bffam=\twelvebf
  \scriptfont\bffam=\tenbf
  \scriptscriptfont\bffam=\eightbf
  \def\bf{\fam\bffam\twelvebf}%
\textfont\scriptfam=\twelvescr
  \scriptfont\scriptfam=\tenscr
  \scriptscriptfont\scriptfam=\eightscr
  \def\script{\fam\scriptfam\twelvescr}%
\textfont\msbfam=\twelvemsb
  \scriptfont\msbfam=\tenmsb
  \scriptscriptfont\msbfam=\eightmsb
  \def\bb{\fam\msbfam\twelvemsb}%
\textfont\truecmr=\twelvetruecmr
  \scriptfont\truecmr=\tentruecmr
  \scriptscriptfont\truecmr=\eighttruecmr
  \def\truerm{\fam\truecmr\twelvetruecmr}%
\textfont\truecmsy=\twelvetruecmsy
  \scriptfont\truecmsy=\tentruecmsy
  \scriptscriptfont\truecmsy=\eighttruecmsy
\tt \ttglue=.5em plus.25em minus.15em 
\setbox\strutbox=\hbox{\vrule height7pt depth2pt width0pt}%
\normalbaselineskip=15pt
\normalbaselines
\rm
}
%
\fontdimen16\tensy=2.7pt
\fontdimen13\tensy=4.3pt
\fontdimen17\tensy=2.7pt
\fontdimen14\tensy=4.3pt
\fontdimen18\tensy=4.3pt
\fontdimen16\eightsy=2.7pt
\fontdimen13\eightsy=4.3pt
\fontdimen17\eightsy=2.7pt
\fontdimen14\eightsy=4.3pt
\fontdimen18\eightsy=4.3pt
%
\def\hexnumber#1{\ifcase#1 0\or1\or2\or3\or4\or5\or6\or7\or8\or9\or
 A\or B\or C\or D\or E\or F\fi}
\mathcode`\=="3\hexnumber\truecmr3D
\mathchardef\not="3\hexnumber\truecmsy36
\mathcode`\+="2\hexnumber\truecmr2B
\mathcode`\(="4\hexnumber\truecmr28
\mathcode`\)="5\hexnumber\truecmr29
\mathcode`\!="5\hexnumber\truecmr21
\mathcode`\(="4\hexnumber\truecmr28
\mathcode`\)="5\hexnumber\truecmr29

\def\tilde{\mathaccent"0\hexnumber\truecmr7E }
\def\bar{\mathaccent"0\hexnumber\truecmr16 }

\def\dot{\mathaccent"0\hexnumber\truecmr5F }
\def\Phi{\mathchar"0\hexnumber\truecmr08 }
\def\Gamma {\mathchar"0\hexnumber\truecmr00 }
\def\Delta {\mathchar"0\hexnumber\truecmr01 }
\def\Theta {\mathchar"0\hexnumber\truecmr02 }
\def\Lambda{\mathchar"0\hexnumber\truecmr03 }
\def\Xi {\mathchar"0\hexnumber\truecmr04 }
\def\Pi{\mathchar"0\hexnumber\truecmr05 }
\def\Sigma{\mathchar"0\hexnumber\truecmr06 }
\def\Upsilon {\mathchar"0\hexnumber\truecmr07 }
\def\Phi {\mathchar"0\hexnumber\truecmr08 }
\def\Psi {\mathchar"0\hexnumber\truecmr09 }
\def\Omega{\mathchar"0\hexnumber\truecmr0A }
\newcount\EQNcount \EQNcount=1
\newcount\CLAIMcount \CLAIMcount=1
\newcount\SECTIONcount \SECTIONcount=0
\newcount\SUBSECTIONcount \SUBSECTIONcount=1
\def\ifff(#1,#2,#3){\ifundefined{#1#2}%
\expandafter\xdef\csname #1#2\endcsname{#3}\else%
\immediate\write16{!!!!!doubly defined #1,#2}\fi}
\def\NEWDEF #1,#2,#3 {\ifff({#1},{#2},{#3})}
\def\actualnumber{\number\SECTIONcount}
\def\EQ(#1){\lmargin(#1)\eqno\tag(#1)}
\def\NR(#1){&\lmargin(#1)\tag(#1)\cr}  
\def\tag(#1){\lmargin(#1)({\rm \actualnumber}.\number\EQNcount)
 \NEWDEF e,#1,(\actualnumber.\number\EQNcount)
\global\advance\EQNcount by 1
}
\def\SECT(#1)#2\par{\lmargin(#1)\SECTION#2\par
\NEWDEF s,#1,{\actualnumber}
}
\def\SUBSECT(#1)#2\par{\lmargin(#1)
\SUBSECTION#2\par 
\NEWDEF s,#1,{\actualnumber.\number\SUBSECTIONcount}
}
\def\CLAIM #1(#2) #3\par{
\vskip.1in\medbreak\noindent
{\lmargin(#2)\bf #1\ \actualnumber.\number\CLAIMcount.} {\sl #3}\par
\NEWDEF c,#2,{#1\ \actualnumber.\number\CLAIMcount}
\global\advance\CLAIMcount by 1
\ifdim\lastskip<\medskipamount
\removelastskip\penalty55\medskip\fi}
\def\CLAIMNONR #1(#2) #3\par{
\vskip.1in\medbreak\noindent
{\lmargin(#2)\bf #1.} {\sl #3}\par
\NEWDEF c,#2,{#1}
\global\advance\CLAIMcount by 1
\ifdim\lastskip<\medskipamount
\removelastskip\penalty55\medskip\fi}
\def\SECTION#1\par{\vskip0pt plus.3\vsize\penalty-75
    \vskip0pt plus -.3\vsize
    \global\advance\SECTIONcount by 1
    \beforesectionskip\noindent
{\sectionsize\sectiontype \actualnumber.\ #1}
    \EQNcount=1
    \CLAIMcount=1
    \SUBSECTIONcount=1
    \nobreak\sectionskip\noindent}
\def\SECTIONNONR#1\par{\vskip0pt plus.3\vsize\penalty-75
    \vskip0pt plus -.3\vsize
    \global\advance\SECTIONcount by 1
    \beforesectionskip\noindent
{\sectionsize\sectiontype  #1}
     \EQNcount=1
     \CLAIMcount=1
     \SUBSECTIONcount=1
     \nobreak\sectionskip\noindent}
\def\SUBSECTION#1\par{\vskip0pt plus.2\vsize\penalty-75%
    \vskip0pt plus -.2\vsize%
    \beforesectionskip\noindent%
{\subsectionsize\subsectiontype \actualnumber.\number\SUBSECTIONcount.\ #1}
    \global\advance\SUBSECTIONcount by 1
    \nobreak\sectionskip\noindent}
\def\SUBSECTIONNONR#1\par{\vskip0pt plus.2\vsize\penalty-75
    \vskip0pt plus -.2\vsize
\beforesectionskip\noindent
{\subsectionsize\subsectiontype #1}
    \nobreak\sectionskip\noindent\noindent}
\def\ifundefined#1{\expandafter\ifx\csname#1\endcsname\relax}
\def\equ(#1){\ifundefined{e#1}$\spadesuit$#1\else\csname e#1\endcsname\fi}
\def\clm(#1){\ifundefined{c#1}$\spadesuit$#1\else\csname c#1\endcsname\fi}
\def\sec(#1){\ifundefined{s#1}$\spadesuit$#1
\else Section \csname s#1\endcsname\fi}
\let\endarg=\par
\def\finish{\def\endarg{\par\endgroup}}
\def\start{\endarg\begingroup}

 \def\beginFROM{\start\parskip=0pt\vskip\baselineskip
\def\finish{\def\endarg{\egroup\par\endgroup}}
  \vbox\bgroup\obeylines\eightpoint\em\finish}

\def\ABSTRACT#1\par{
\vskip 1in {\noindent\sectionsize\sectiontype Abstract.} #1 \par}

\def\TODAY{\number\day~\ifcase\month\or January \or February \or March \or
April \or May \or June
\or July \or August \or September \or October \or November \or December \fi
\number\year\timecount=\number\time
\divide\timecount by 60
}
\newcount\timecount
\def\DRAFT{\def\lmargin(##1){\strut\vadjust{\kern-\strutdepth
\vtop to \strutdepth{
\baselineskip\strutdepth\vss\rlap{\kern-1.2 truecm\eightpoint{##1}}}}}
\font\footfont=cmti7
\footline={{\footfont \hfil File:\jobname, \TODAY,  \number\timecount h}}
}
\newbox\strutboxJPE
\setbox\strutboxJPE=\hbox{\strut}
\def\subitem#1#2\par{\vskip\baselineskip\vskip-\ht\strutboxJPE{\item{#1}#2}}
\gdef\strutdepth{\dp\strutbox}
\def\lmargin(#1){}
\def\period{\unskip.\spacefactor3000 { }}
%
%
\newbox\noboxJPE
\newbox\byboxJPE
\newbox\paperboxJPE
\newbox\yrboxJPE
\newbox\jourboxJPE
\newbox\pagesboxJPE
\newbox\volboxJPE
\newbox\preprintboxJPE
\newbox\toappearboxJPE
\newbox\bookboxJPE
\newbox\bybookboxJPE
\newbox\publisherboxJPE
\newbox\inprintboxJPE
\def\refclearJPE{
   \setbox\noboxJPE=\null             \gdef\isnoJPE{F}
   \setbox\byboxJPE=\null             \gdef\isbyJPE{F}
   \setbox\paperboxJPE=\null          \gdef\ispaperJPE{F}
   \setbox\yrboxJPE=\null             \gdef\isyrJPE{F}
   \setbox\jourboxJPE=\null           \gdef\isjourJPE{F}
   \setbox\pagesboxJPE=\null          \gdef\ispagesJPE{F}
   \setbox\volboxJPE=\null            \gdef\isvolJPE{F}
   \setbox\preprintboxJPE=\null       \gdef\ispreprintJPE{F}
   \setbox\toappearboxJPE=\null       \gdef\istoappearJPE{F}
   \setbox\inprintboxJPE=\null        \gdef\isinprintJPE{F}
   \setbox\bookboxJPE=\null           \gdef\isbookJPE{F}  \gdef\isinbookJPE{F}
     
   \setbox\bybookboxJPE=\null         \gdef\isbybookJPE{F}
   \setbox\publisherboxJPE=\null      \gdef\ispublisherJPE{F}
     
}

\def\ref{\refclearJPE\bgroup}
\def\no   {\egroup\gdef\isnoJPE{T}\setbox\noboxJPE=\hbox\bgroup}
\def\by   {\egroup\gdef\isbyJPE{T}\setbox\byboxJPE=\hbox\bgroup}
\def\paper{\egroup\gdef\ispaperJPE{T}\setbox\paperboxJPE=\hbox\bgroup}
\def\yr{\egroup\gdef\isyrJPE{T}\setbox\yrboxJPE=\hbox\bgroup}
\def\jour{\egroup\gdef\isjourJPE{T}\setbox\jourboxJPE=\hbox\bgroup}
\def\pages{\egroup\gdef\ispagesJPE{T}\setbox\pagesboxJPE=\hbox\bgroup}
\def\vol{\egroup\gdef\isvolJPE{T}\setbox\volboxJPE=\hbox\bgroup\bf}
\def\preprint{\egroup\gdef
\ispreprintJPE{T}\setbox\preprintboxJPE=\hbox\bgroup}
\def\toappear{\egroup\gdef
\istoappearJPE{T}\setbox\toappearboxJPE=\hbox\bgroup}
\def\inprint{\egroup\gdef
\isinprintJPE{T}\setbox\inprintboxJPE=\hbox\bgroup}
\def\book{\egroup\gdef\isbookJPE{T}\setbox\bookboxJPE=\hbox\bgroup\em}
\def\publisher{\egroup\gdef
\ispublisherJPE{T}\setbox\publisherboxJPE=\hbox\bgroup}
\def\inbook{\egroup\gdef\isinbookJPE{T}\setbox\bookboxJPE=\hbox\bgroup\em}
\def\bybook{\egroup\gdef\isbybookJPE{T}\setbox\bybookboxJPE=\hbox\bgroup}
\newdimen\refindent
\refindent=5em
\def\endref{\egroup \sfcode`.=1000
 \if T\isnoJPE
 \hangindent\refindent\hangafter=1
      \noindent\hbox to\refindent{[\unhbox\noboxJPE\unskip]\hss}\ignorespaces
     \else  \noindent    \fi
 \if T\isbyJPE    \unhbox\byboxJPE\unskip: \fi
 \if T\ispaperJPE \unhbox\paperboxJPE\unskip\period \fi
 \if T\isbookJPE {\it\unhbox\bookboxJPE\unskip}\if T\ispublisherJPE, \else.
\fi\fi
 \if T\isinbookJPE In {\it\unhbox\bookboxJPE\unskip}\if T\isbybookJPE,
\else\period \fi\fi
 \if T\isbybookJPE  (\unhbox\bybookboxJPE\unskip)\period \fi
 \if T\ispublisherJPE \unhbox\publisherboxJPE\unskip \if T\isjourJPE, \else\if
T\isyrJPE \  \else\period \fi\fi\fi
 \if T\istoappearJPE (To appear)\period \fi
 \if T\ispreprintJPE Pre\-print\period \fi
 \if T\isjourJPE    \unhbox\jourboxJPE\unskip\ \fi
 \if T\isvolJPE     \unhbox\volboxJPE\unskip\if T\ispagesJPE, \else\ \fi\fi
 \if T\ispagesJPE   \unhbox\pagesboxJPE\unskip\  \fi
 \if T\isyrJPE      (\unhbox\yrboxJPE\unskip)\period \fi
 \if T\isinprintJPE (in print)\period \fi
\filbreak
}
\def\hexnumber#1{\ifcase#1 0\or1\or2\or3\or4\or5\or6\or7\or8\or9\or
 A\or B\or C\or D\or E\or F\fi}
\textfont\msbfam=\tenmsb
\scriptfont\msbfam=\sevenmsb
\scriptscriptfont\msbfam=\fivemsb
\mathchardef\varkappa="0\hexnumber\msbfam7B
\newcount\FIGUREcount \FIGUREcount=0
\newdimen\figcenter
\def\fig(#1){\ifundefined{fig#1}%
\global\advance\FIGUREcount by 1%
\NEWDEF fig,#1,{Fig.\ \number\FIGUREcount}
\immediate\write16{ FIG \number\FIGUREcount : #1}
\fi
\csname fig#1\endcsname\relax}
\def\figure #1 #2 #3 #4\cr{\null%
\ifundefined{fig#1}%
\global\advance\FIGUREcount by 1%
\NEWDEF fig,#1,{Fig.\ \number\FIGUREcount}
\immediate\write16{  FIG \number\FIGUREcount : #1}
\fi
{\goodbreak\figcenter=\hsize\relax
\advance\figcenter by -#3truecm
\divide\figcenter by 2
\midinsert\vskip #2truecm\noindent\hskip\figcenter
\includegraphics{#1}\vskip 0.8truecm\noindent \vbox{\eightpoint\noindent
{\bf\fig(#1)}: #4}\endinsert}}
\def\figurewithtex #1 #2 #3 #4 #5\cr{\null%
\ifundefined{fig#1}%
\global\advance\FIGUREcount by 1%
\NEWDEF fig,#1,{Fig.\ \number\FIGUREcount}
\immediate\write16{ FIG \number\FIGUREcount: #1}
\fi
{\goodbreak\figcenter=\hsize\relax
\advance\figcenter by -#4truecm
\divide\figcenter by 2
\midinsert\vskip #3truecm\noindent\hskip\figcenter
\includegraphics{#1}{\hskip\texpscorrection\input #2 }\vskip 0.8truecm\noindent \vbox{\eightpoint\noindent
{\bf\fig(#1)}: #5}\endinsert}}
\def\figurewithtexplus #1 #2 #3 #4 #5 #6\cr{\null%
\ifundefined{fig#1}%
\global\advance\FIGUREcount by 1%
\NEWDEF fig,#1,{Fig.\ \number\FIGUREcount}
\immediate\write16{ FIG \number\FIGUREcount: #1}
\fi
{\goodbreak\figcenter=\hsize\relax
\advance\figcenter by -#4truecm
\divide\figcenter by 2
\midinsert\vskip #3truecm\noindent\hskip\figcenter
\includegraphics{#1}{\hskip\texpscorrection\input #2 }\vskip #5truecm\noindent \vbox{\eightpoint\noindent
{\bf\fig(#1)}: #6}\endinsert}}
\catcode`@=11
\def\footnote#1{\let\@sf\empty 
  \ifhmode\edef\@sf{\spacefactor\the\spacefactor}\/\fi
  #1\@sf\vfootnote{#1}}
\def\vfootnote#1{\insert\footins\bgroup\eightpoint
  \interlinepenalty\interfootnotelinepenalty
  \splittopskip\ht\strutbox 
  \splitmaxdepth\dp\strutbox \floatingpenalty\@MM
  \leftskip\z@skip \rightskip\z@skip \spaceskip\z@skip \xspaceskip\z@skip
  \textindent{#1}\footstrut\futurelet\next\fo@t}
\def\fo@t{\ifcat\bgroup\noexpand\next \let\next\f@@t
  \else\let\next\f@t\fi \next}
\def\f@@t{\bgroup\aftergroup\@foot\let\next}
\def\f@t#1{#1\@foot}
\def\@foot{\strut\egroup}
\def\footstrut{\vbox to\splittopskip{}}
\skip\footins=\bigskipamount 
\count\footins=1000 
\dimen\footins=8in 
\catcode`@=12 

\def\CC{{\script C}}

\def\LL{{\script L}}

\def\OO{{\script O}}

\def\HALF{{\textstyle{1\over 2}}}

\def\QED{\hfill\smallskip
         \line{$\hfill{\vcenter{\vbox{\hrule height 0.2pt
	\hbox{\vrule width 0.2pt height 1.8ex \kern 1.8ex
		\vrule width 0.2pt}
	\hrule height 0.2pt}}}$
               \ \ \ \ \ \ }
         \bigskip}
\def\real{{\bf R}}

\def\integer{{\bf Z}}
\def\Re{{\rm Re\,}}

\def\PROOF{\medskip\noindent{\bf Proof.\ }}
\def\REMARK{\medskip\noindent{\bf Remark.\ }}
\def\LIKEREMARK#1{\medskip\noindent{\bf #1.\ }}
\tenpoint
\normalbaselineskip=5.25mm
\baselineskip=5.25mm
\parskip=10pt
\beforesectionskipamount=24pt plus8pt minus8pt
\sectionskipamount=3pt plus1pt minus1pt
\def\em{\it}
\normalbaselineskip=12pt
\baselineskip=12pt
\parskip=0pt
\parindent=22.222pt
\beforesectionskipamount=24pt plus0pt minus6pt
\sectionskipamount=7pt plus3pt minus0pt
\overfullrule=0pt
\hfuzz=2pt
\nopagenumbers
\headline={\ifnum\pageno>1 {\hss\tenrm-\ \folio\ -\hss} \else
{\hfill}\fi}
\if T\islinuxolivetti
\font\titlefont=cmbx10 scaled\magstep2

\font\toplinefont=cmr10
\font\pagenumberfont=cmr10
\let\tenpoint=\rm
\else
\font\titlefont=ptmb at 14 pt

\font\toplinefont=cmcsc10
\font\pagenumberfont=ptmb at 10pt
\fi
\newdimen\itemindent\itemindent=1.5em

\def\textindent#1{\indent\llap{#1\enspace}\ignorespaces}
\def\item{\par\noindent
\hangindent\itemindent\hangafter=1\relax
\setitemmark}
\def\setitemindent#1{\setbox0=\hbox{\ignorespaces#1\unskip\enspace}%
\itemindent=\wd0\relax
\message{|\string\setitemindent: Mark width modified to hold
         |`\string#1' plus an \string\enspace\space gap. }%
}
\def\setitemmark#1{\checkitemmark{#1}%
\hbox to\itemindent{\hss#1\enspace}\ignorespaces}
\def\checkitemmark#1{\setbox0=\hbox{\enspace#1}%
\ifdim\wd0>\itemindent
   \message{|\string\item: Your mark `\string#1' is too wide. }%
\fi}
\def\SECTION#1\par{\vskip0pt plus.2\vsize\penalty-75
    \vskip0pt plus -.2\vsize
    \global\advance\SECTIONcount by 1
    \beforesectionskip\noindent
{\sectionsize\sectiontype \actualnumber.\ #1}
    \EQNcount=1
    \CLAIMcount=1
    \SUBSECTIONcount=1
    \nobreak\sectionskip\noindent}

\headline
{\ifnum\pageno>1 {\toplinefont Cahn-Hilliard Equation}
\hfill{\pagenumberfont\folio}\fi}\null
\let\truett=\tt
\fontdimen3\tentt=2pt\fontdimen4\tentt=2pt
\def\tt{\hfill\break\null\kern -3truecm\truett
 }
\def\CH{Cahn-Hilliard }\def\L{{\rm L}}
\def\d{{\rm d}}
\def\D{{\rm D}}
\parskip0pt\parindent20pt
\let\epsilon=\varepsilon
\let\theta=\vartheta
\let\kappa=\varkappa
\let\phi=\varphi
\setitemindent{iii)}
\def\CC{{\cal C}}
{\titlefont{\centerline{Non-Linear Stability Analysis of 
 }}}
\vskip 0.5truecm
{\titlefont{\centerline { Higher Order Dissipative Partial
Differential Equations}}
\vskip 0.5truecm
{\it{\centerline{J.-P. Eckmann${}^{1,2}$ and C.E. Wayne${}^3$}}
\vskip 0.3truecm
{\eightpoint
\centerline{${}^1$D\'ept.~de Physique Th\'eorique, Universit\'e de Gen\`eve,
CH-1211 Gen\`eve 4, Switzerland}
\centerline{${}^2$Section de Math\'ematiques, Universit\'e de Gen\`eve,
CH-1211 Gen\`eve 4, Switzerland}
\centerline{${}^3$Dept.~of Mathematics, Boston University,
111 Cummington St., Boston, MA 02215, USA}
}}
\vskip 0.5truecm
{\eightpoint\narrower\baselineskip 11pt
\LIKEREMARK{Abstract} We extend the invariant manifold method for analyzing the
asymptotics of dissipative partial differential equations on
unbounded spatial domains to treat equations in which the linear 
part has order greater than two.  One important example of this
type of equation which we analyze in some detail is the Cahn-Hilliard
equation.  We analyze the marginally stable solutions of this
equation in some detail.  A second context in which such equations
arise is in the Ginzburg-Landau equation, or other pattern forming
equations, near a codimension-two bifurcation.  
}
\vfill\eject
\tenpoint
\def\tt{\hfill\break\null\kern -3truecm\truett 
\#\#\#\#\#\#\#\#\#\#\#\#\#\#  }
\SECTION Introduction and statement of results

In this paper, we extend the methods developed in [W1], [W2], [EWW],
to study the asymptotic behavior of marginally stable non-linear
PDE's. These are PDE's such as
$$
\partial_t u \,=\, P(-i\nabla_x) u + W'(u)~,
$$
where $u=u(x,t)$, with $x\in \real^d$, and where $P$ is a
polynomial. In the papers cited above, we have treated essentially
parabolic problems, {\it i.e.}, the case where $P(\xi)=-\xi^2$. In this paper, we
extend the problem to non-parabolic cases such as $P(\xi)=-\xi^4$,
where $P(-i\nabla_x)$ has continuous spectrum all the way up to 0. We
deal in particular with the stability analysis of the \CH equation [CH]
in an infinite domain.
Where appropriate, we indicate how to formulate the assumptions for
more general differential operators and non-linearities.  

The \CH
equation models the dynamics of a material with the following 3
properties: {\parskip=0pt
\item{i)}The material prefers one of two concentrations that can
coexist at a given temperature.
\item{ii)}The material prefers to be spatially uniform.
\item{iii)}The total mass is conserved.
}
\medskip
The first point above means that we should consider a potential with 2
minima with equal critical values, and for concreteness, we will
choose $W(u)=(1-u^2)^2$.\footnote{${}^*$}{In our example, the
curvatures of the two minima are equal. This does not seem to be
necessary for our proofs.}  The \CH equation is then
$$
\partial_t u\,=\, \Delta \bigl (-\Delta u  + W'(u)\bigr)~,
\EQ(CH1)
$$
or, expanding,
$$
\partial_t u\,=\, -\Delta^2 u - 4 \Delta u + 4 \Delta u^3~.
\EQ(CH2)
$$
We will be interested specifically in the {\em non-linear stability}
of the spatially uniform states, $u(x,t)\equiv u_0$.

It is obvious that constants are solutions of Eq.\equ(CH2), for any
$u_0$. Furthermore, it is easy to check that these solutions are
(locally) linearly stable for $|u_0|>3^{-1/2}$, and linearly unstable
for $|u_0|<3^{-1/2}$.  We concentrate our analysis on the remaining
case, namely $u_0=\pm 3^{-1/2}$. In this case, linearizing about
$u_0=3^{-1/2}$ leads to the linear equation
$$
\partial_t v \,=\, -\Delta ^2 v~,
\EQ(lch)
$$
which has spectrum in $(-\infty,0]$ and corresponds to the case
$P(\xi)=-\xi^4$. It is obvious that bounded
initial data lead to solutions which tend to 0 as $t\to\infty$ and the
purpose of this paper is to study under which conditions the addition
of the nonlinear terms does {\em not} change the stability of the
solutions. This is more difficult, for two reasons: First, as we have
said, the spectrum of the linearized problem extends all the way to 0,
and second, the nonlinearity does not have a sign. 

Another, more complicated, example
of a similar nature is provided by time-independent
solutions of the Ginzburg-Landau equations (on $\real$)
$$
\partial_t u\,=\, \partial_x^2 u +u - u  |u|^2~,
\EQ(GL1)
$$
which are exactly on the borderline between being Eckhaus stable and
Eckhaus unstable. These solutions are
$$
u_q(x)\,=\, e^{iqx}\sqrt{1-q^2} ~,
$$
with $q=1/\sqrt{3}$, {\it cf.} [EG].
We will not prove that this problem scales like the Cahn-Hilliard
equations, but only describe a program which we believe
would lead to a proof. The first part of the analysis
of this problem would follow rather
closely that given in [EEW] for the Swift-Hohenberg equation. Letting
$u^*= u_q $ for the critical value $q=1/\sqrt{3}$, and writing
$u=u^*+v$, the equation for $v$ is
$$
\partial_t v \,=\, \partial_x^2 v + v - 2 v |u^*|^2 -\bar v (u^*)^2
+\OO(v^2)~.
\EQ(GL2)
$$
It has a linear part which is like a
Schr\"odinger operator in a periodic potential (the inhomogeneity
$u^*$). 
This can be handled by
going to Floquet variables, namely
setting
$$
v(x,t)\,=\,\int _{-q}^q \d k\, e^{ikx}v_k(x,t)~,
$$
where $v_k$ is $\pi/q$-periodic in $x$:
$$
v_k(x,t)\,=\,\sum_{m\in\integer} e^{2imqx} v_{k,m}(t)~.
$$
The linear part of Eq.\equ(GL2) leaves the subspace spanned by the
$v_k$ invariant, and has discrete spectrum in each such subspace.
The spectrum is in $\sigma\le0$ and the largest eigenvalue is
$-\OO(k^4)$ when $q $ equals its critical value $q=1/\sqrt{3}$ (which
is the case we discuss here).
In this sense, the problem of the marginal Eckhaus instability
resembles the problem of the Cahn-Hilliard equation.
At this point, the discussion of the problem follows the
techniques we developed in [EWW]. We would like to rescale as we will
do below for the Cahn-Hilliard equation and its generalizations, but
the problem will be more complicated because the Brillouin zone is
restricted to $k\in [-q,q]$. We then have to check that the
non-linearity is ``irrelevant'' in the terminology developed
below. Again, as in [EWW], we believe that this will not be quite the
case, 
but the saving grace will be that the projection of the potentially
non-irrelevant modes onto the eigenstates corresponding to the
$-\OO(k^4)$ term
vanish to some higher degrees because of translation invariance of the
original problem, {\it cf.} [EWW, Section 4], and [S].

We place our examples in the following more general setting. Consider
equations of the form
$$
{\partial u\over \partial t}\,=\, (-1)^{n+1} \Delta^n u + F(u,
\{\partial_x^\alpha u\})~,
\EQ(X1)
$$
where the multi-indices $\alpha $ satisfy $|\alpha |\le 2n-1$, and
$x\in \real^d$, $t\ge1$. Furthermore, $F$ is a polynomial in $u$ and
its derivatives.  We wish to study the asymptotics of the solution $u$
of \equ(X1) as $t\to\infty $. First, one introduces scaling variables
by defining
$$
u(x,t)\,=\,{1\over t^{ d/(2n)} }\, v\bigl ( {x\over t^{ 1/(2n)}} ,
\log t\bigr )~.
\EQ(X2)
$$
Introducing new variables $\xi=x/t^{1/(2n)}$ and $\tau = \log(t
+t_0)$, with $t_0$ an arbitrary positive constant, the Eq.\equ(X1) is
transformed to the non-autonomous problem
$$
{\partial v\over \partial \tau }\,=\, (-1)^{n+1} \Delta_\xi^n v +
{1\over 2n} \xi\cdot \nabla_\xi v + {d\over 2n} v + e^{({2n+d\over
2n})\tau} F(e^{-{d\tau \over 2n}}v, \{e^{-({|\alpha| +d\over 2n})\tau
} \partial_\xi^\alpha v\})~.
\EQ(X3)
$$
The analysis of this equation involves two steps:
\item{i)}An analysis of the linear operator
\item{ii)}A determination of which non-linear terms are relevant.
\medskip
As we will see, the term $1/(2n) \xi\cdot \nabla_\xi$ plays an
important r\^ole in the analysis of this linear operator as it allows
us to push the continuous spectrum of the operator more and more into
the stable region by working in Sobolev spaces with higher and higher
polynomial weights. These weights force the functions to decrease more
and more rapidly near $|x|=\infty$.  Taking Fourier transforms on both
sides of Eq.\equ(X3) we obtain:
$$
{\partial \tilde v\over \partial \tau }\,=\, - (p\cdot p)^{n}\tilde v -
{1\over 2n} p\cdot \nabla_p \tilde v+ e^{({2n+d\over 2n})\tau}
F^*(e^{-{d\tau \over 2n}}\tilde v, \{e^{-({|\alpha| +d\over 2n})\tau }
(-ip)^\alpha \tilde v\})~,
\EQ(X4)
$$
where $F^*$ is the polynomial $F$, written in terms of convolution
products, (see the discussion of the non-linearities below).

We will discuss the form of the non-linear terms below, and consider
first the linear operator
$$
\LL \,=\, -(p\cdot p)^{n} -{1\over 2n} p\cdot \nabla _p~.
\EQ(X5)
$$
A straightforward calculation shows that $\LL $ has the countable
set of eigenvalues
$$
\lambda _j\,=\, -{ j\over 2n}~,\quad j=0,1,2,\dots~,
\EQ(X6)
$$
with eigenfunctions (written in multi-index notation),
$$
\phi_ \alpha (p)\,=\, p^{\alpha }e^{-(p\cdot p)^{n}}~,
\EQ(af1)
$$
and $|\alpha |=j$.

If we consider $\LL $ as acting on the Sobolev spaces
$$
\tilde{H}_{\ell,m}\,=\,\left \{ \tilde v ~:~
\|p^\alpha \partial_p^\beta \tilde v\|_{\L^2} <\infty ~,
{\rm ~for~all~} |\alpha |\le \ell, |\beta |\le m\right \}~,
$$
then $\LL $ will have continuous spectrum in the half-plane $\Re
\lambda < -\sigma_m$ in addition to the
eigenvalues above. 
By choosing $m$ appropriately, we can force this
continuous spectrum arbitrarily far into the left half-plane, and the
dominant behavior of the linear operator will be dictated by the
eigenvalues with the largest real part.

\REMARK In order to switch back and forth from the Fourier
transform representation of $\LL$ to the un-Fourier transformed
representation of this operator with ease, we also consider
the Sobolev spaces
$$
{H}_{\ell,m}\,=\,\left \{  v ~:~
\|\partial_x^\alpha x^\beta  v\|_{\L^2} <\infty ~,
{\rm ~for~all~} |\alpha |\le \ell, |\beta |\le m\right \}~.
$$
Note that Fourier transformation is an isomorphism from
$\tilde{H}_{\ell,m}$ to ${H}_{\ell,m}$.

Note that $\LL$ is {\em not} sectorial, and therefore we know of no
way to bound the
semi-group generated by $\LL$ by spectral
information alone. However, in Appendix A, we develop an integral
representation of the semi-group and we then show that it satisfies
the estimates needed for the invariant manifold theorem.

We next discuss which terms in the non-linearity are ``relevant.''
Consider a monomial
$$
A\,=\,
\prod _{j=0}^s \left ( \partial_x^{\alpha ^{(j)}} u\right  )^{k_j}~,
\EQ(X7)
$$
where the $\alpha ^{(j)}$ are distinct multi-indices. After rescaling
and taking Fourier transforms this becomes
$$
\eqalign{
\tilde A\,&=\,
\exp\left(({2n+d\over 2n})\tau \right)\,\,\exp\left (-\sum_{j=0}^s 
\bigl ({|\alpha ^{(j)}|+d\over 2n}\bigr )k_j\tau \right )\cr
&\times
\bigl ((-ip)^{\alpha ^{(0)}}\tilde v\bigr )^{*k_0}*\cdots*
\bigl ((-ip)^{\alpha ^{(s)}}\tilde v\bigr )^{*k_s}~.\cr
}
\EQ(X8)
$$
Here, $*$ denotes the convolution product. If we combine the powers of
$\tau $ in the exponential, we see that if
$$
2n+d\,<\,\sum_{j=0}^s
\bigl ({|\alpha ^{(j)}|+d}\bigr )k_j~, 
\EQ(X9)
$$
then the coefficient of this term will go to zero exponentially fast
in $\tau $, and hence it will be irrelevant from the point of view of
the long time behavior of the solutions.

\LIKEREMARK{Definitions}A monomial like \equ(X8) is called {\em
irrelevant} if it satisfies the inequality \equ(X9). It is called {\em
critical} if the l.h.s.~of Eq.\equ(X9) is equal to the r.h.s, and {\em
relevant} in the remaining case.

These definitions are suggested by the following
which is our first main result:
\CLAIM Theorem(T1) Assume all terms in the non-linearity in Eq.\equ(X1)
are irrelevant. For any solution $u(x,t)$ of Eq.\equ(X1) with
sufficiently small initial conditions in $H_{m,\ell}$ (with $\ell>
(2n-1)+d/2$ and $m>(n+2)/(2n)$ ), there is a constant $B^*$, depending
on the initial conditions, such that for every $\epsilon >0$,
$$
\lim_{t\to\infty }
t^{({d+1\over 2n}-\epsilon )} \left \| u(x,t)-{B^*\over t^{d/(2n)}}
f^*({x\over t^{1/(2n)}}) \right \|_{\L^\infty } \,=\,0~.
$$
Here,
$$
f^*(\xi)\,=\,{1\over (2\pi)^{d/2}} \int\, \d^dp\, e^{ip\cdot \xi }
e^{-(p\cdot p)^n}~.
\EQ(Xfstar)
$$

\REMARK This theorem is a special case of a more detailed analysis
which will be given below. That analysis will allow us to compute, in
principle, the {\em form} of the solutions of Eq.\equ(X1) up to
$\OO(t^{-k})$, for any $k>0$. We note that if one only wanted the
first order asymptotics of the solution, one could also use the
renormalization group analysis of [BKL].

We now apply the \clm(T1) to the Cahn-Hilliard equation. Writing $u=
3^{-1/2} + w$, the function $w$ is seen to satisfy
$$
{\partial w\over \partial t}\,=\,-\Delta^2 w +\sqrt{3} \Delta (w^2) +
\Delta(w^3)~.
\EQ(X10)
$$
Upon expanding $\Delta(w^2)$ we obtain two types of terms---those of
the form $w(\partial_{x_i}^2 w)$ and those of the form
$(\partial_{x_i}w)^2$.  In both cases,
$$
\sum (|\alpha ^{(j)}|+d )k_j \,=\, 2d+2~.
$$
Since $n=2$ in this example, these terms will be irrelevant if
$4+d<2d+2$, that is in dimensions $d>2$. Also, the term $\Delta(w^3)$
is irrelevant for $d>1$.  Thus, as a corollary to \clm(T1) we get
immediately
\CLAIM Corollary(T2) Solutions of the Cahn-Hilliard equation in
dimension $d\ge3$, with initial conditions sufficiently close (in $H_{1,1}$) to
the constant solution $u\equiv 3^{-1/2}$ behave asymptotically as
$$
u(x,t)\,=\, {1\over 3^{1/2}} + {B^*\over t^{d/4}} f^*\bigl ({x\over
t^{1/4}}\bigr ) + \OO\bigl ( {1\over t^{(d+1)/4 -\epsilon }}\bigr )~.
\EQ(Xcor2)
$$

\REMARK We will examine below what happens in the cases $d=1,2$. The
case $d=2$ is of particular interest because its non-linearity is
critical in the renormalization group terminology.

\SECTION Invariant manifolds

Note that spectral subspaces corresponding to eigenvalues of $\LL$ 
are automatically invariant manifolds for the semi-flow defined by
the linear part of Eq.\equ(X4). The aim of this section is to
demonstrate that the full non-linear problem has similar invariant
manifolds in a neighborhood of the origin. This then shows that the
conceptual understanding of what is happening can be gained purely
from a knowledge of $\LL $, (and the scaling behavior of the
non-linearity).

We begin with a proposition concerning the linear semi-group generated
by $\LL $.
\CLAIM Proposition(T3) Let $P_k$ denote the projection onto the
spectral subspace associated with the eigenvalues $\bigl \{ {-j\over
2n}\bigr \}_{j=0}^k$, and let $Q_k=P_k^\perp$ (in $H_{\ell,m}$).  If
$m>(n+k+1)/(2n)$, then there exists $C_k>0$  such that
the semi-group generated by $\LL $ satisfies
$$
\eqalign{
\|Q_k e^{\tau \LL } Q_k v\|_{{\ell,m}}\,&\le\,{C_k\over t^{q/2n}
} \exp\bigl ( {k+1\over 2n}\tau \bigr )\|v\|_{{\ell-q,m}}~,
 q=0,1,\dots,2n-1~.\cr}
\EQ(X12)
$$

\PROOF The proof, which is presented in Appendix A, is modeled on the
proof in [EWW] which treats the case $n=1$.

Given such estimates on the linear evolution, the construction of
invariant manifolds is straightforward.  Denote by $y$ the coordinates
on the (finite-dimensional) range of $P_k$, and let $z=Q_k\tilde
v$. Finally let $\eta= e^{-\tau /(2n)}=(t+t_0)^{-1/(2n)}$.
Then, applying the projection operators $P_k$ and $Q_k$ to
Eq.\equ(X4), it can be written as the system of equations
$$
\eqalign{
\dot y\,&=\, \Lambda_k y + f(y,z,\eta)~,\cr
\dot z\,&=\,Q_k \LL  z + g (y,z,\eta)~,\cr
\dot \eta \,&=\, -{\textstyle{1\over 2n}} \eta~,\cr
}
\EQ(X13)
$$
where $\dot{\hphantom X}$ denotes differentiation w.r.t.~$\tau$.  We
next need a bound on the non-linearity:
\CLAIM Lemma(T4) Assume $u\in H_{m,\ell}$ with $\ell> 2n-1+d/2$, and
assume
$$
2n+d \,\le\, \sum_{j=0}^s |{\alpha ^{(j)}+d} |k_j~.
$$
Then the non-linear term Eq.\equ(X8) has $H_{m,\ell-2n+1} $ norm
bounded by
$$
\eta^p \prod _{j=0}^s \|u\|_{{m,\ell-2n+1}}\,=\,
\eta^p \|u\|_{{m,\ell-2n+1}}^K~,
$$
with $p=\sum_{j=0}^s |{\alpha ^{(j)}+d}|k_j - (2n+d)$ and
$K=\sum_{j=0}^s k_j$.

\PROOF Taking the inverse Fourier transform of Eq.\equ(X8), and
substituting $\eta= e^{-\tau/(2n)}$, Eq.\equ(X8) becomes
$$
\eta^p \prod_{j=0}^s \left ( \partial_\xi^{\alpha ^{(j)}}v\right )^{k_j}~.
$$
The result then follows from the Sobolev embedding theorem. Note that
the lemma has the immediate corollary (because $F$ is a polynomial):
\CLAIM Corollary(T5) Under the hypotheses of \clm(T4), for every $r\ge
1$, the non-linear term in Eq.\equ(X4) is a $\CC^r$ function from
$\real\times H_{m,\ell}$ to $H_{m,\ell-2n+1}$.

This corollary in turn implies that the terms in Eq.\equ(X8) and
Eq.\equ(X13)  are all
$\CC^r$ functions. This, in conjunction with the estimates on the
linear semi-group is sufficient to establish the following
\CLAIM Theorem(T6) Suppose that $\ell> 2n-1 +d/2$ and
$m>(n+k+1)/(2n)$. Suppose further that all terms in the nonlinearity
satisfy
$$
2n+d \,\le\, \sum_{j=0}^s |{\alpha ^{(j)}+d} |k_j~.
\EQ(X99)
$$
Then there exists a $\CC^{1+\alpha }$ function $h(\eta,y)$, with
$\alpha >0$, defined in some neighborhood of the origin in
$\real\times\real^{{\rm dim~range}(P_k)}$, such that the manifold
$z=h(\eta,y)$ is left invariant by the semi-flow of
Eq.\equ(X13). Furthermore, any solution of Eq.\equ(X13) which remains
near the origin for all times approaches a solution of
\equ(X13)---restricted to the invariant manifold---at a rate $\OO\bigl(
e^{{k+1-\epsilon \over 2n}\tau}\bigr)$.

\PROOF The existence of the invariant manifold, given the assumptions
on the linear semi-group and the non-linearity, seems, to our
knowledge, not to be 
explicitly spelled out in the literature. The formulation which comes
closest to our needs is the one given in [H], where the assumptions on
the non-linearity are those we have in our case, but the semi-group is
supposed to be analytic. However, Henry's construction of the
invariant manifold  only uses certain bounds on the decay of
the semi-group, and not the stronger assumption of analyticity.
Those  bounds {\em are} true in our case, by \clm(T4). Thus, existence
of the invariant manifold follows in fact from Henry's proof.

Once one knows that the manifold exists, it is also easy
to show that any solution near the origin must approach a solution on
the invariant manifold (see, {\it e.g.}~[C]). Note that even though
our non-linearity is quite smooth, we cannot hope, in general, to
obtain an invariant manifold whose smoothness is greater than
$\CC^{1+\alpha }$, since this smoothness is related to the gap between
the spectrum of $\Lambda_k$, and that of $Q_k \LL  Q_k$, (see,
{\it e.g.}~[LW]).
\SECTION Applications

Here, we show how the existence of the invariant manifold implies
\clm(T1) and related results. To prove \clm(T1), we assume that all
terms in the non-linearity are irrelevant. This means that
Eq.\equ(X99) holds. Suppose further that $k=1$ and that $\ell>
2n-1+d/2$ and $m>(n+2)/(2n)$. These hypotheses guarantee that \clm(T6)
applies and hence any solution near the origin must approach a
solution on the invariant manifold, at a rate $\OO\bigl(
e^{{2-\epsilon \over 2n}\tau }\bigr)$ in $H_{m,\ell}$.

The equations on the invariant manifold can be written as a system of
ordinary differential equations:
$$
\eqalign{
\dot y_0\,&=\, \bigl\langle \phi_ 0^* | f\bigl(
y,h(\eta,y),\eta\bigr)\bigr\rangle~,\cr
\dot y_{1,j}\,&=\, -{\textstyle{1\over 2n}} y_{1,j}+\bigl\langle \phi_ {1,j}^* | f\bigl(
y,h(\eta,y),\eta\bigr)\bigr\rangle~,\quad j=1,\dots,d~,\cr
\dot \eta \,&=\, -{\textstyle{1\over 2n}} \eta~,\cr 
}
\EQ(X14)
$$
where $\phi_0^*$ and $\phi_{1,j}^*$ are the projections onto the
spectral subspace of $\lambda _0$ and $\lambda _1=-1/(2n)$,
respectively. Note that $\lambda _1$ has a $d$-dimensional spectral
subspace.

The important observation to make at this point is that since the
non-linearity is assumed to be irrelevant, there exist constants $C_0$
and $C_1$ such that
$$
\left |\bigl\langle \phi_ {0}^* | f\bigl(
y,h(\eta,y),\eta\bigr)\bigr\rangle\right |\,\le\,C_0 \eta^p~,\quad
\left |\bigl\langle \phi_ {1}^* | f\bigl(
y,h(\eta,y),\eta\bigr)\bigr\rangle\right |\,\le\,C_1\eta^p~,
$$
for some $p\ge1$. Since $\eta(\tau )=e^{-\tau /(2n)}\eta(0)$, this
implies immediately that solutions of Eq.\equ(X14) behave as
$$
\eqalign{
y_0(\tau )\,&=\,B^*+\OO(e^{-\tau /(2n)})~,\cr y_{1,j}(\tau
)\,&=\,\OO(e^{-\tau /(2n)})~.\cr }
$$
The eigenfunction with eigenvalue 0 of $\LL $ is $e^{-p\cdot p}$, or
taking inverse Fourier transform, $f^*$, {\it
cf.}~Eq.\equ(Xfstar). Thus, in $H_{m,\ell}$ solutions of Eq.\equ(X4)
behave as
$$
\tilde v(p,\tau )\,=\, B^*e^{-p\cdot p} +\OO(e^{-\tau /(2n)})~.
$$
Reverting from scaling variables to the unscaled variables $u(x,t)$
and using the Sobolev lemma to estimate the $\L^\infty $ norm in terms
of the $H_{m,\ell}$ norm, we obtain \clm(T1).  Since we observed above
that the non-linearity in the Cahn-Hilliard equation is irrelevant
when $d\ge3$, we immediately see in this case that Eq.\equ(Xcor2)
holds for initial conditions which are close to $u\equiv 3^{-1/2}$,
which yields \clm(T2).

\SECTION The critical case 

We now consider the Cahn-Hilliard equation in dimension $d=2$, which
is the critical case in terms of the renormalization group terminology
[BKL]. This means that in some non-linear terms the inequality
Eq.\equ(X9) becomes an equality.

In the Cahn-Hilliard equation, when $d=2$ (and $n=2$), we see that the
quadratic term is critical, and the cubic term is irrelevant.
Note that \clm(T6) still implies the existence of an invariant
manifold tangent at the origin to the eigenspace of $\lambda_0$.  This
means that when written in the form of Eq.\equ(X13), the non-linearity
can be written as the sum of 2 pieces---one quadratic in $y$ and $z$
which is independent of $\eta$ (and hence critical) and a cubic piece
in $y$ and $z$ which is linear in $\eta$ (and hence irrelevant). This
implies that the Eqs.\equ(X14), when reduced to the invariant
manifold, take the form
$$
\eqalign{
\dot y_0\,&=\,\bigl\langle \phi_ 0^* | f^{(2)}\bigl(
y,h(\eta,y),\eta\bigr)\bigr\rangle +\bigl\langle \phi_ 0^* |
f^{(3)}\bigl( y,h(\eta,y),\eta\bigr)\bigr\rangle~,\cr
\dot y_{1,j}\,&=\, -{\textstyle{1\over 4}} y_{1,j}+\bigl\langle \phi_
{1,j}^* | f^{(2)}\bigl(
y,h(\eta,y),\eta\bigr)\bigr\rangle+\bigl\langle \phi_ {1,j}^* |
f^{(3)}\bigl( y,h(\eta,y),\eta\bigr)\bigr\rangle~,\quad j=1,2~,\cr
\dot \eta \,&=\, -{\textstyle{1\over 4}} \eta~.\cr 
}\EQ(X17)
$$

We now exploit the form of the non-linear term in Eq.\equ(X10), namely
$3^{1/2}\Delta(w^2)+\Delta(w^3)$, plus the fact that the eigenfunction
$\phi_0^*\equiv 1$. Thus if we integrate by parts, we find that
$$
\bigl\langle \phi_ 0^* | f^{(2)}\bigl(
y,h(\eta,y),\eta\bigr)\bigr\rangle +\bigl\langle \phi_ 0^* |
f^{(3)}\bigl( y,h(\eta,y),\eta\bigr)\bigr\rangle\,=\,0~,
$$
so that in Eq.\equ(X17), $\dot y_0\equiv0$ and thus
$y_0(t)=y_0(0)$. This reflects the fact that the Cahn-Hilliard
equation conserves mass.

Since from Eq.\equ(X17) we also see that $y_{1,j}=\OO(e^{-\tau /4})$,
we find upon reverting to the unscaled variables the second main result:
\CLAIM Theorem(T7) For $d=2$, if the initial conditions of the
Cahn-Hilliard equation are sufficiently close in $H_{1,4}$ to the
stationary state $u\equiv 3^{-1/2}$, then the solution behaves
asymptotically as
$$
u(x,t)\,=\,{1\over 3^{1/2}}+{B^*\over t^{1/2}}f^*\bigl( {x\over
t^{1/4}}\bigr) +\OO\bigl( {1\over t^{3/4-\epsilon }}\bigr)~.
$$

\SECTION The relevant case

Here, we consider the case of $d=1$ where one term of the
non-linearity is relevant. This necessitates a change of strategy,
because the quadratic term is proportional to $\eta^{-1}$ and hence
the non-linear terms in Eq.\equ(X13) are not smooth enough to apply
the invariant manifold theorem. In order to circumvent this
difficulty, we choose a scaling different from Eq.\equ(X2). Consider
again the Cahn-Hilliard equation, Eq.\equ(X10), with $u=3^{-1/2}+w$.
In $d=1$, we get
$$
{\partial w\over \partial t}\,=\, - {\partial^4\over \partial x^4} w
+3^{1/2} {\partial^2\over \partial x^2}\bigl( w^2\bigr) +
{\partial^2\over \partial x^2}\bigl( w^3\bigr)~.
\EQ(X18)
$$
Now let $w(x,t)=t^{-1/2} W(x/t^{1/4},\log t)$.
Then $W$ satisfies
$$
{\partial W\over \partial \tau }\,=\, -\partial_\xi^4 W + {1\over 4}
\xi \cdot\partial_\xi W +{1\over 2} W + 3^{1/2} \partial_\xi^2 \bigl(
W^2\bigr) +e^{-\tau/2} \partial_\xi^2\bigl( W^3\bigr)~.
\EQ(X19)
$$
Proceeding as in the other cases, we define the linear operator
$L_1=-\partial_\xi^4 +{1\over 4} \xi\partial_\xi + \HALF$, which in
Fourier variables becomes
$$
\LL _1\,=\,-p^4 -{\textstyle{1\over 4}} p \partial_p + {\textstyle{1\over 4}}~,
$$
so that it has eigenvalues $\mu_m={1-j\over 4}$, $j=0,1,\dots$~.
Thus, unlike the operator $\LL$, we have one eigenvalue lying in the
right half-plane.  Let $\tilde \eta=e^{-\tau /8}$, and let
$y_0$ and $y_1$ denote the amplitudes of the eigenvectors with
eigenvalues $\mu_0$ and $\mu_1$. Then Eq.\equ(X19) takes the form
$$
\eqalign{
\dot y_0\,&=\,{\textstyle {1\over 4}} y_0+ f_0(y_0,y_1,\eta,y^\perp)~,\cr
\dot y_1\,&=\,f_1(y_0,y_1,\eta,y^\perp)~,\cr
\dot \eta~\, \,&=\, -{\textstyle{1\over 8}} \eta~,\cr 
\dot y^\perp \,&=\, QL_1 y^\perp + f^\perp(y_0,y_1,\eta,y^\perp)~.\cr 
}
\EQ(X21)
$$
Here, $Q$ is the projection onto the complement of the eigenspaces
corresponding to $\mu_0$ and $\mu_1$, $y^\perp =QW$, and $f_0$, $f_1$,
and $f^\perp$ are the projections of the non-linearity onto the
various subspaces.

Since the spectrum of $QL_1Q $ lies in the half-plane $\Re \mu\le
{1\over 4}$, we can construct an invariant manifold for Eq.\equ(X21)
which is the graph of a function $h^\perp(y_0,y_1,\eta)$, and every
solution of Eq.\equ(X21) which remains in a neighborhood of the origin
will approach this manifold at a rate $\OO(e^{-\tau /4})$. What is
more, the equations on the invariant manifold are extremely simple in
this case, since the projections onto the ``0'' and ``1'' components
correspond to integrating with respect to the functions 1 and $x$,
respectively. Applying these projections to the non-linearity and
integrating once, resp.~twice by parts, we see that these projections
of the non-linear terms vanish. Thus, the equations on the invariant
manifold of Eq.\equ(X21) are simply
$$
\dot y_0\,=\,{\textstyle{1\over 4}} y_0~,
\quad
\dot y_1\,=\,0~,\quad \dot {\tilde \eta} = -
{\textstyle{1\over 8}} \tilde \eta~.
$$
Thus, {\em as long as the solution of Eq.\equ(X21) remains in a
neighborhood of the origin}, it will be of the form
$$
\eqalign{
y_0(\tau)\,&=\,e^{\tau /4} y_0(0)~,\cr y_1(\tau)\,&=\,y_1(0)~,\cr
\tilde \eta(\tau) \,&=\,e^{-\tau /8}\tilde \eta(0)~,\cr 
y^\perp(\tau)\,&=\,\OO(e^{-\tau /4})~.\cr }
\EQ(X23)
$$
Thus, we see that the solution either leaves the neighborhood of the
origin,
or its asymptotic behavior can be read off from Eq.\equ(X23).
Note that the solutions that remain near the origin must have
$y_0=0$.  Thus:
\CLAIM Theorem(T8) Suppose that the initial condition of the
Cahn-Hilliard equation is of the form $u_0=3^{-1/2}+w_0$ with $w_0$
small in the $H_{m,\ell} $ norm for some $\ell\ge4$ and
$m\ge2$. Assume furthermore that $\int_{-\infty }^\infty \d
x\,w_0(x)=0$.  Then the solution is of the form
$$
u(x,t)\,=\, {1\over 3^{1/2}} + {B^{**}\over t^{1/2}} f^{**} \bigl(
{x\over t^{1/4}}\bigr) +\OO\bigl( {1\over t^{3/4-\epsilon }}\bigr)~,
$$
where
$$
B^{**} \,=\,\int_{-\infty }^\infty \d x\,xw_0(x)~,
\quad
f^{**}(\xi)\,=\,{1\over \sqrt{2\pi}}\int_{-\infty }^\infty \d p\,
e^{ip\xi}e^{-p^4}~.
$$

\REMARK The constant $B^*$
in \clm(T1) is not as easy to describe because there, the
non-linearity in the equation for $y_0$ did {\em not}
necessarily disappear.

\PROOF The proof is an obvious modification of the one of \clm(T1),
taking into account the special form of the eigenfunctions
corresponding to the eigenvalues $\mu_0$ and $\mu_1$.
\def\actualnumber{A}
\SECTIONNONR Appendix. Bounds on the linear semi-group

In this appendix, we sketch the proof of \clm(T3). The proof is quite
similar to the estimates on the linear semi-group in Appendix B of
[EWW], (which was given for the case of a one-dimensional  Laplacian,
or in the
present notation $n=d=1$) so we concentrate
only on the points where the present argument differs from the one in
[EWW].

We begin with the representation
$$
\bigl (e^{\tau L}\bigr )(x)\,=\, {e^{\tau d\over 2n}\over 2\pi^d }
\int \d^dz\, g(z,\tau ) v\bigl (e^{\tau \over 2n}(x+z)\bigr )~,
\EQ(b1)
$$
where
$$
g(z,\tau )\,=\,\int \d^d k \,e^{i k\cdot z} \exp\bigl ({-(k\cdot k)^n
}(1-e^{-\tau })\bigr )~.
\EQ(b1g)
$$
As in [EWW], the action of the semi-group is analyzed by considering
separately the behavior of the part far from the origin and that
close to the origin. The new difficulty here is that we do not have an
explicit representation of $g$ as in the case $n=1$. However, the
technique of estimating the long-time
behavior will remain essentially the same.  Let
$\chi_R$ be a smooth characteristic function which vanishes for
$|x|<R$ and is equal to 1 for $|x|> 4R/3$. We start by studying the
region far from the origin. The analog of Proposition B.2 of [EWW] is

\CLAIM Proposition(more1) For every $\ell \ge 1$ and every $m\ge0$, there
exist a $\gamma>0$ and a $C(\ell,m)<\infty $ such that for all $v\in
H_{\ell,m}$ one has
$$
\left \| \chi_R e^{\tau L} v\right \|_{\ell,m}\,\le\,
{C(\ell,m)\over \bigl (a(\tau )\bigr )^{q\over 2n}} 
e^{({{\tau}\over{2n}})(d+\ell)} \left
(e^{-\tau m/2}+e^{-\gamma R^{2n/(2n-1)}}\right )\, \|v\| _{\ell-q,m}~,
\EQ(equmore1)
$$
for $q=0,1,\dots, 2n-1$. Here, $a(\tau )=1-e^{-\tau }$.

The crucial step in proving this estimate is to derive the asymptotics
of $g(z,\tau )$ for large $z$. This will replace the explicit
(Gaussian) estimates for the $d=1$, $n=1$ case analyzed in [EWW]. This
estimate is provided by the following
\CLAIM Proposition(more2) The kernel $g(z,\tau )$ decays faster than
any inverse power of $z$ for $|z|$ large. In fact, one has the
estimate 
$$
|g(z,\tau)|\,\le\,
C a(\tau )^{-{d\over 2n}} 
\exp\bigl (
-\gamma (|z|^{2n}/a(\tau ))^{1\over 2n-1}
\bigr )~,
\EQ(b1a)
$$
for some $\gamma = \gamma (n,d)>0$.

\REMARK If $n=1$, we recover the explicit bound on the Green's
function: 
$$
{C\over \sqrt{a(\tau )}} e^{-\gamma (2,d) z^2/a(\tau )}~.
$$

\PROOF We need to estimate the quantity
$$
I_{n,d}\,=\,\int \d^d k\, e^{-a(\tau )\bigl
 (\sum_{j=1}^d k_j^2\bigr )^n}  e^{i \sum
_{j=1}^d k_j x_j} ~.
\EQ(b1b)
$$
By rotational symmetry, it suffices to bound the preceding
expression for $x$ lying on the positive real axis.  Setting
$x = 2n a(\tau)  z^{2n-1}$, and $k = (p,q)$,
with $p \in \real$, and $q \in \real^{d-1}$, 
this means that we must bound
$$
X\,=\,\int \d p \,\d^{d-1}q \exp\bigl ( -a(\tau) (p^2 + q \cdot q)^n +
 2inp a(\tau)  z^{2n-1}
\bigr )~.
$$
If we rescale the variables as
$p=z t$, and $q=z s$, then we have
$$
X\,=\,z^d \int \d t \,\d^{d-1} s \exp\bigl (
 - a(\tau) z^{2n}\bigl ((t^2 + s \cdot s)^n
+ 2int\bigr )\bigr )~.
$$
\REMARK Note that the polynomial $(t^2 + s \cdot s)^n + 2int$ is
independent of $z$.

We will bound $X$ by taking advantage of the fact that the integrand
is an entire function and translate the contour of integration so that
it passes through at least one critical point of the exponent.  These critical
points occur at $s=0$ and the roots of $t^{2n-1} =i$ -- {\it i.e.},
at the points $t_k = \exp(i{{\pi(4k+1)}\over{2(2n-1)}})$, 
$k=0,1,2, \dots, 2n-2$.

Inserting this expression into the exponent of the integrand of $X$,
we see that the value of the polynomial at the critical points
is
$$
\eqalign{
-a(\tau) &z^{2n}\left (\exp( i{{2n \pi(4k+1)}\over{2(2n-1)}})
 -2ni\exp( i{{\pi(4k+1)}\over{2(2n-1)}})\right )\cr
\,&=\,
(2n-1)a(\tau) z^{2n}\exp(i{{\pi(4k+2n)}\over{2(2n-1)}})~.\cr
}
$$
In particular, if we take $k=0$, then the real part of the critical
value is
$$
(2n-1) a(\tau) z^{2n} \cos\bigl (\pi/2+\pi/(4n-2)\bigr )
\,\approx\, - a(\tau) z^{2n} \cdot {{\pi}\over{2}}~,
$$
when $n$ is large (and is negative for all $n>0$).
Integrating over the region $\real+t_0$ and observing that there is
only one critical point on this line, we get, using the techniques of
H\"ormander:
$$
X\,\approx\, a(\tau)^{d/(2n)} e^{-C_n a(\tau)  z^{2n}}~,
$$
with $C>0$ and $C_n\to \pi/2$ as $n\to\infty $, when $z\to \infty $.
Reverting to the original variables, this leads to
$$
I_{n,d}\,\approx\, a(\tau )^{d/(2n)} 
e^{-D_n x^{2n/(2n-1)}/a(\tau)^{1/(2n-1)}}~,
$$
where $D_n=C_n / (2n)^{1/(2n-1)}$.

We now consider the action of the semi-group on functions localized
inside a ball of radius $R$. A key observation here is the following
lemma.
Let $\phi_ 0(x)$ denote the eigenfunction of $L$ with eigenvalue 0 and
let $T(x)=\tilde \phi_ 0(x)^{1/2}$ (note that $\tilde \phi_ 0(x)>0$
for all $x$).  
\CLAIM Lemma(af1) The operator $H=T^{-1} L T$ is self-adjoint on
(a dense domain in)
$\L^2(\real^d)$ and has the same eigenvalues as $L$.

\PROOF The proof is a straightforward calculation.
We note further that if $\phi_\alpha $ are the eigenfunctions of $L$
then the eigenfunctions of $H$ are $\psi_\alpha (x)=\bigl
(T^{-1}\phi_{\alpha}\bigr )(x)$.

If we take the inverse Fourier transform of the eigenfunctions $\phi_
\alpha (p)$ of Eq.\equ(af1), we see that
$$
|\tilde \phi_ \alpha (x)|\,\approx\,
C|x|^{|\alpha |} \exp( -\gamma |x|^{2n\over 2n-1})~,
$$
for some $\gamma>0$, using the same sort of estimates as those used to
bound the kernel $g$ of the semi-group.
Thus, for $|x|$ sufficiently large, we get
$$
|\tilde \psi_ \alpha (x)|\,\approx\,
C|x|^{|\alpha |} \exp( -\HALF\gamma |x|^{2n\over 2n-1})~.
$$

The usefulness of introducing the operator $H$ is that it is
sectorial, since it is self-adjoint and bounded below.
Therefore, the associated semi-group can be estimated from
spectral information alone. In particular, if $P_k$ denotes the
projection onto the spectral subspace spanned by the eigenfunctions
with eigenvalues $0$, $-{1\over 2n}$, ${-2\over 2n}$,\dots, ${-k\over
2n}$, and $Q_k$ is defined by $Q_k=1-P_k$, then we have
a bound on the operator norm of $Q_k e^{\tau H} Q_k$

$$
\|Q_k e^{\tau H} Q_k\|\,\le\,
C_k e^{-\tau k/(2n)}~.
\EQ(af2)
$$
We can use this information to bound the semi-group associated with
$L$. Note that if we denote by $P_k^{(0)}$ and $Q_k^{(0)}$ the projection associated
with the spectral subspaces of $L$ (as we did for $H$), the we have
the identity:
$$
e^{\tau L} Q_k^{(0)}v\,=\,
e^{\tau L} Q_k^{(0)}(1-\chi_R)v+e^{\tau L} Q_k^{(0)}\chi_Rv~.
$$
Since 
$\chi_R v$ is localized away from the origin, it can be studied
with the help of \clm(more1), so we only focus on the other term. And
there we get
$$
\eqalign{
\| e^{\tau L} Q_k^{(0)}(1- \chi_R) v\|_{q,r}
\,&=\,
\| T T^{-1} e^{\tau L}T T^{-1}Q_k^{(0)}
T T^{-1}(1- \chi_R) v\|_{q,r}
\cr
\,&=\,\bigl  \| T\bigl (e^{\tau H} Q_k \bigr )\bigl (T^{-1}(1- \chi_R)
v\bigr )\bigr \|_{q,r} \cr
\,& \le\,
C \exp\bigl ({-\tau {k+1\over 2n}}\bigr )\bigl \|
T^{-1} (1- \chi_R)
v\bigr \|_{q,r} ~.\cr
}
$$
Using now the information that $|T^{-1}(x)|\le C \exp( \gamma
|x|^{2n\over 2n-1})$ and that $(1- \chi_Rv)(x)=0$ when $|x|> 4R/3$, we get
$$
\left \|
T^{-1}(1- \chi_R) v \right \|_{q,r}\,\le\, C \exp\bigl (
\gamma (4R/3)^{2n\over 2n-1}\bigr )\|v \|_{q,r} ~,
$$
so that finally
$$
\left \|
e^{\tau L} Q_k^{(0)}(1- \chi_R) v \right \|_{\ell,m}\,\le\, C \exp\bigl (
\gamma (4R/3)^{2n\over 2n-1}\bigr )
\exp \bigl ( -\tau  {k+1\over 2n}\bigr )\|v \|_{\ell,m} ~.
$$

Thus we have proven:
\CLAIM Proposition(smallsup) Under the hypotheses of\/ \clm(more1),
there exist constants $C(\ell,m) >0$ and $\gamma >0$, such
that for all $v\in H_{\ell,m}$, one has
$$
\left \|
e^{\tau L} Q_k^{(0)}(1- \chi_R) v \right \|_{\ell,m}\,\le\, C \exp\bigl (
\gamma (4R/3)^{2n\over 2n-1}\bigr )
\exp \bigl ( -\tau  {k+1\over 2n}\bigr )\|v \|_{\ell,m} ~.
$$

We now return to the:
\LIKEREMARK{Proof of \clm(T3)}As in [EWW] it is only necessary
to consider the term with highest derivative in $\|\chi_R e^{\tau
\LL}v\|_{\ell,m}$. All other terms are easier to estimate. Also, as in
that paper, we use the fact that 
$$
\D^{\ell} e^{\tau \LL}\,=\,e^{\tau \ell/(2n)} e^{\tau \LL} \D^{\ell}~,
\EQ(b10)
$$
where $\D^\ell$ is a shorthand notation for a product of derivatives
w.r.t.~the $x_j$ of total degree $\ell$.
Thus, 
$$
\bigl ( e^{\tau \LL} \D^{\ell} v\bigr )(x)\,=\,
{e^{\tau d/(2n)}\over (2\pi)^d}
\int \d^dz\,
g(z,\tau ) \bigl (\D^{\ell} v\bigr )(e^{\tau /(2n)}(x+z))~.
\EQ(b11)
$$
First consider the $q=0$ case of \equ(equmore1). Then
$$
\|\chi_R e^{\tau
\LL}v\|_{\ell,m}
\,\le\,
{e^{\tau d/(2n)}\over (2\pi)^d}
\int \d^dz\,
|g(z,\tau )| \bigl \|w^m \chi_R\bigl (\D^{\ell} v\bigr )(e^{\tau
/(2n)}(.+z))\bigr \|_2~,
\EQ(b12)
$$
where $w$ is the operator of multiplication by $(1+x\cdot x)^{1/2}$.
Note that the conclusions of Lemma B.4 of [EWW] do not depend on the
exact form of $g$ and so it also holds in the present situation and we
have
\CLAIM Lemma(b3) One has the bounds
$$
\bigl \|w^r\chi_Rv(e^{\tau
/(2n)}(.+z))\bigr \|_2^2
\,\le\,
\cases
{
Ce^{-\tau m} \|v\|_{0,m}^2~, & if $|z|\le 7R/8$~,\cr
C(1+|z|^2)^r \|v\|_{0,m}^2,& if $|z|> 7R/8$~.
\cr
}
$$

\REMARK Note that the proof in [EWW] is also unaffected by the
dimension $d$ in which we work.

Now use \clm(b3) to bound the integral in \equ(b12) by writing it as
an integral over $|z|\le7R/8$ and an integral over $|z|>7R/8$.
The integral over $|z|\le7R/8$ is bounded with the aid of \clm(b3) as
$$
C {e^{\tau d/(2n)}\over (2\pi)^d}
\int_{|z|\le 7R/8}
|g(z,\tau )| e^{-\tau m/2}
\|v\|_{\ell,m}
\,\le\,
C(n) e^{\tau d/2n} e^{-\tau m} \|v\|_{\ell,m}~,
\EQ(b14)
$$
where the last step used the estimates of \clm(T7) to show that
$\int \d z |g(z,\tau )|\le C$, with $C$ independent of $\tau $.

To estimate the integral in the outer region, we use the second part
of \clm(b3) and then bound it by
$$
C {e^{\tau d/(2n)}\over (2\pi)^d}
\int_{|z| > 7R/8}
|g(z,\tau )| e^{-\tau r/2}
\|v\|_{\ell,m}
\,\le\,
C(n,m) e^{\tau d/(2n)} \exp\bigl (-\gamma R^{2n\over 2n-1}\bigr )
\|v\|_{\ell,m}~,
\EQ(b15)
$$
for some $\gamma >0$, where, again, we have used the estimates of
decay in \clm(T7) both to extract the factor of 
$ \exp\bigl (-\gamma R^{2n\over 2n-1}\bigr )$ as well as to bound
the integral over $z$. Combining Eqs.\equ(b1), \equ(b14), and
\equ(b15), we get the $q=0$ case of Eq.\equ(equmore1).

We next indicate how to treat the $q>0$ cases of Eq.\equ(equmore1). Consider
the case $q=1$. We can rewrite Eq.\equ(b11) by integrating by parts
once w.r.t.~one component of $z$, for example $z_1$.
Then,
$$
\bigl (
e^{\tau \LL} \D^{\ell} v
\bigr )(x) \,=\,
{e^{\tau d/(2n)}e^{-\tau /(2n)}\over (2\pi)^d}
\int \d^d z \bigl (\D_{z_1} g\bigr )(z,\tau)
\bigl (\D^{\ell-1}v\bigr )(e^{\tau /(2n)}(x+z))
~.
\EQ(b16)
$$
Differentiating \equ(b1g) w.r.t.~$z_1$ gives
$$
(\D_{z_1} g\bigr )(z,\tau)
\,=\,
\int \d^d k \,ik_1\,\exp(iq\cdot z) \exp\bigl (
(k\cdot k)^n (1-e^{-\tau }) \bigr )~.
\EQ(b17)
$$
To estimate \equ(b17), first replace $k$ by $ p_j=a(\tau )^{1/(2n)}
k_j$, where, as before $a(\tau )=1-e^{-\tau }$.
Then,
$$
(\D_{z_1} g\bigr )(z,\tau)
\,=\,
{i\over a(\tau )^{d/(2n)}} {1\over  a(\tau )^{1/(2n)}}
\int
\d^dp\,
p_1 \exp \bigl ((p\cdot p)^n \bigr )
\exp \bigl (i p\cdot z / a(\tau )^{1/(2n)}\bigr )
~.
\EQ(b18)
$$
The estimate of the integral in Eq.\equ(b18) now follows as before,
since the extra factor of $p_1$ does not cause any trouble as it is
easily offset by the exponentially decaying terms.

One now uses the Schwarz inequality to rewrite
$$
\|
\chi_R
e^{\tau \LL}
\D^\ell v 
\|_{\ell,m}
\,\le\,
{ e^{\tau {d/(2n)}}
\over 
(2\pi)^d}
\int \d^d z\,
\left | \D_{z_1} g(z,\tau )\right |
\,\left \|
w^m \chi_R \bigl ( \D^{\ell-1} v\bigr )
(e^{\tau /2}(.+z)) 
\right \|_2~,
\EQ(b19)
$$
and then proceeds as in the case when $q=0$, breaking the integral
over $z$ into the same two pieces as before.
These two pieces are then estimated with the aid of \clm(b3). Note
that while the factor $a(\tau )^{-d/(2n)}$ of Eq.\equ(b18) will be
absorbed when one integrates w.r.t~$z$, the remaining factor of
$a(\tau )^{-1/(2n)}$ will remain in the final bound of Eq.\equ(equmore1).
The bounds for  $q=2,3,\dots\,2n-1$ follow in a similar fashion.

To complete the proof of \clm(T3), first rewrite
$$
e^{\tau \LL} Q_k\,=\,
e^{\tau \LL/2} Q_k e^{\tau \LL/2}
\,=\,
e^{\tau \LL/2} Q_k \chi_R e^{\tau \LL/2} +
e^{\tau \LL/2} Q_k(1-\chi_R) e^{\tau \LL/2}~.
\EQ(191)
$$
The second of these terms involves an estimate of the action of $
e^{\tau \LL/2} Q_k$ on a function localized near the origin, so by
\clm(smallsup), we get a bound
$$
\|e^{\tau \LL/2} Q_k(1-\chi_R)e^{\tau \LL/2} v \|_{\ell,m}
\,\le\,
C_q \exp\left (
\gamma  \bigl ({4R\over 3}\bigr )^{2n/( 2n-1)}
-\HALF {k+1\over 2n}\right )
\|v\|_{\ell-q,m}~.
\EQ(192)
$$
We use \clm(more1) to bound the first term of \equ(191):
$$
{C(\ell,m)\over a(\tau )^{q\over (2n)}}
e^{\tau \ell/2}
\left (
e^{-{1\over 4} \tau m} + e^{-\gamma R^{2n/ (2n-1)}}
\right )
\|v\|_{\ell-q,m}~.
\EQ(193)
$$
As a preliminary step, we note that if we first choose
$r$ and $R$ such that
$$e^{\tau q/2}
\left (
e^{-{1\over 4} \tau r} + e^{-\gamma R^{2n/ (2n-1)}}
\right )\,\le\,
e^{-\mu ((k+1)/(2n))}~,
$$
then for sufficiently small $\mu$ (roughly speaking $\mu\sim \HALF
\bigl (1+(4/3)^{2n/(2n-1)}\bigr )^{-1}$), the Eqs.\equ(192) and
\equ(193) imply
$$
\|e^{\tau \LL}Q_k\|\,\le\,
{C\over a(\tau)^{q/(2n)}}
e^{-\mu {k+1\over 2n}}
\|v\|_{\ell-q,k}
\,=\,
{C\over a(\tau)^{q/(2n)}}
e^{-\mu |\lambda_{k+1}|}
\|v\|_{\ell-q,k}~.
$$

This shows that the projection of the semi-group onto the complement
of the eigenspace spanned by the first $k$ eigenvalues decays with a
rate proportional to the eigenvalue $\lambda _{k+1}$. We can sharpen
the decay rate so that we obtain a rate like $\exp\bigl (-(1-\epsilon
)|\lambda _{k+1}|\bigr )$ by the techniques of [EWW], (see
Eq. B.14 and following) and this completes the proof of \clm(T3)

\SECTIONNONR References

{\eightpoint{
\setitemindent{[EWW]}
\ref 
\no BKL
\by Bricmont, J., A. Kupiainen, and G. Lin
\paper Renormalization group and asymptotics of solutions of nonlinear
parabolic equations
\jour Comm. Pure Appl. Math.
\vol 47
\pages 893--922
\yr 1994
\endref
\ref
\no EG
\by Eckmann, J.-P., and Th. Gallay
\paper Front solutions of the Ginzburg-Landau equation
\jour Comm. Math. Phys.
\vol 152
\pages 221--248
\yr 1993
\endref
\ref
\no C
\by J. Carr
\book The Centre Manifold Theorem and its Applications
\publisher Springer-Verlag
\yr 1983
\endref
\ref
\no CH
\by Cahn, J. W., and H. E. Hilliard
\paper Free energy of a nonuniform system I. Interfacial free energy
\jour J. Chem. Phys.
\vol 28
\pages 258--267
\yr 1958
\endref
\ref
\no EWW
\by Eckmann, J.-P.,  C.E. Wayne, and P. Wittwer
\paper Geometric stability analysis for periodic solutions
                   of the {S}wift-{H}ohenberg equation
\jour Comm. Math. Phys.
\vol 190
\pages 173--211
\yr 1997
\endref
\ref
\no H
\by Henry, D.
\book Geometric Theory of Semilinear Parabolic Equations
\publisher Lecture Notes in Mathematics, {\bf  840}, Springer, Berlin
\yr 1981
\endref
\ref
\no LW
\by Llave, R., and C.E. Wayne
\paper On {I}rwin's proof of the pseudostable manifold theorem
\jour Math. Z.
\vol 219
\pages 301--321
\yr 1995
\endref
\ref 
\no S
\by G. Schneider
\paper Diffusive stability of spatial periodic solutions of the 
		{S}wift-{H}ohenberg equation
\jour Comm. Math. Phys.
\vol 178
\pages 679--702
\yr 1996
\endref
\ref
\no W1
\by Wayne, C.E.
\paper Invariant manifolds for parabolic partial differential
		 equations on unbounded domains
\jour Arch. Rat. Mech. Anal.
\vol 138
\pages 279--306
\yr 1997
\endref
\ref
\no W2
\by Wayne, C.E.
\paper Invariant manifolds and the asymptotics of parabolic
                equations in cylindrical domains {\rm in}
book Proceedings of the first {US}/{C}hina Conference
                   on Differential Equations, Hangzhou PR
\publisher International Press
\yr 1997
\endref
}}}
\bye